\documentclass[english,twocolumn]{svjour3}
\usepackage[T1]{fontenc}
\usepackage[latin9]{inputenc}
\usepackage{url}
\usepackage{amsmath}
\usepackage{graphicx}

\makeatletter
\RequirePackage{fix-cm}

\smartqed  
\usepackage{upquote}

\makeatother

\usepackage{babel}
\begin{document}
\title{Numerical simulation of GUE two-point correlation and cluster functions}
\titlerunning{Two-point eigenvalue correlation functions}
\author{Adam James Sargeant}
\authorrunning{Adam James Sargeant}
\institute{Adam James Sargeant \at Departamento de Ciências Exatas e Aplicadas,
Instituto de Ciências Exatas e Aplicadas, Universidade Federal de
Ouro Preto, Rua Trinta e Seis, 115, Loanda, João Monlevade, Minas
Gerais, 35931-008, Brasil\\
Tel.: +55-31-3852-8709\\
adam@ufop.edu.br}
\date{Received: date / Accepted: date}
\maketitle
\begin{abstract}
Numerical simulations of the two-point eigenvalue correlation and
cluster functions of the Gaussian unitary ensemble (GUE) are carried
out directly from their definitions in terms of deltas functions.
The simulations are compared with analytical results which follow
from three analytical formulas for the two-point GUE cluster function:
(i) Wigner's exact formula in terms of Hermite polynomials, (ii) Brezin
and Zee's approximate formula which is valid for points with small
enough separations and (iii) French, Mello and Pandey's approximate
formula which is valid on average for points with large enough separations.
It is found that the oscillations present in formulas (i) and (ii)
are reproduced by the numerical simulations if the width of the function
used to represent the delta function is small enough and that the
non-oscillating behaviour of formula (iii) is approached as the width
is increased.

\end{abstract}

\section{Introduction\label{sec:Introduction}}

Random matrices were introduced by Wishart in multivariate statistics
in the 1920s \cite{Wishart1928} and Wigner in the study of neutron
resonances in the 1950s \cite{Wigner1957}. Since then random matrix
theory has found further applications in nuclear physics \cite{Porter1965,Brody1981,Mitchell2010,Weidenmueller2009,Sargeant2005,Hussein2016}
as well as in quantum and wave chaos \cite{Bohigas1984,Bohigas1991},
quantum chromodynamics \cite{Stephanov1999}, mesoscopic physics \cite{Beenakker1994,Hussein2017},
quantum gravity \cite{Ambjorn1990}, numerical computation \cite{Edelman2014},
number theory \cite{Hayes2003,Forrester2015,Wolf2020} and complex
systems \cite{Erguen2009}. 

The main objects of analytical studies of correlations of the eigenvalues
of random matrices are typically the correlation functions themselves,
in particular the two-point correlation function, while the main objects
of numerical studies are typically derivative correlation measures
such as spacing distributions and the number variance, which are more
convenient numerically and are simpler to interpret visually \cite{Dyson1962,Dyson1963,Bohigas1991}.
There is however insight to be gained from numerical simulations of
the correlation functions themselves and in this paper we numerically
calculate the GUE two-point correlation and cluster functions directly
from their definitions in terms of delta functions and compare the
numerical calculations with some known analytical formulas. 

The paper is organised as follows: In Section \ref{sec: level den}
the GUE is defined and the level density is discussed. In Section
\ref{sec:Two-point} the correlation and cluster functions are defined
and some known analytical results are listed. In Section \ref{sec: numerical}
numerical simulations of the correlation and cluster functions are
presented and compared with the analytical formulas listed in Section
\ref{sec:Two-point}. In Section \ref{concl} the results are summarised
and some conclusions drawn and in the Appendix some considerations
on coding the numerical simulations are made.

\section{Level density\label{sec: level den}}

In Gaussian random matrix theory, the Hamiltonian is represented by
a square matrix $H$ of independent random variables whose probability
density is given by \cite{Porter1965}

\begin{equation}
P(H)=\left(\frac{A}{\pi}\right)^{\frac{\beta}{4}N(N-1)+\frac{N}{2}}\exp\left(-A\mathrm{Tr}H^{2}\right),\label{probH}
\end{equation}
where $N$ is the matrix size, $\beta$ is the Dyson index (number
of real variables per matrix element) and $A$ characterises the variance
of the matrix elements. The Gaussian orthogonal ensemble (GOE), Gaussian
unitary ensemble (GUE) and Gaussian symplectic ensemble (GSE) are
further defined by three possible symmetries of $H$ which result
in real eigenvalues \cite{Porter1965,Livan2018}. 

In the case of the GUE, a member of the ensemble of Hamiltonians may
be constructed from \cite{Livan2018,Edelman2014}

\begin{equation}
H=\frac{M+M^{\dagger}}{2}\label{eq:H}
\end{equation}
where $M$ is an $N\times N$ complex matrix whose elements have both
real and imaginary parts drawn from a Gaussian distribution with mean
$\mu=0$ and standard deviation $\sigma$. Then

\begin{equation}
\sigma^{2}=\frac{1}{2A}\label{eq:variance}
\end{equation}
is the variance of the diagonal elements of $H$. The variance of
the real and imaginary parts of the off-diagonal matrix elements is
$\sigma^{2}/2$ thanks to the trace in Eq. (\ref{probH}) \cite{Livan2018}.

Denoting the $N$ real eigenvalues of $H$ by $E_{i}$, the eigenvalue
or level density may be written

\begin{equation}
\rho(E)=\sum_{i=1}^{N}\delta(E-E_{i})\label{rho def}
\end{equation}
which is the number of levels per unit energy at energy $E$. The
level density averaged over a large number of realizations of the
ensemble is denoted by $\overline{\rho(E)}$. In the limit that $N\rightarrow\infty$,
$\overline{\rho(E)}$ for Gaussian ensembles is given by the Wigner
semicircle law \cite{Wigner1957}

\begin{equation}
\rho_{W}(E)=\frac{N}{\pi a^{2}/2}\sqrt{a^{2}-E^{2}}\label{semi}
\end{equation}
with the radius of the semicircle given by

\begin{equation}
a=\sigma\sqrt{2\beta N}=\sqrt{\frac{\beta N}{A}},\label{eq:a}
\end{equation}
with the Dyson index $\beta=2$ for the GUE. For finite $N$ the average
level density for the GUE is given by \cite{Wigner1962,Mehta1963,Bronk1964,Mehta1967}

\begin{equation}
\rho_{H}(E)=\exp\Bigl(-\frac{E^{2}}{2\sigma^{2}}\Bigr)\sum_{j=0}^{N-1}\pi_{j}^{2}(E),\label{denH}
\end{equation}
where
\begin{equation}
\pi_{j}(E)=\frac{1}{\left(\sqrt{2\pi}\sigma2^{j}j!\right)^{\frac{1}{2}}}H_{j}\Bigl(\frac{E}{\sqrt{2}\sigma}\Bigr)\label{piOrtho}
\end{equation}
and the $H_{j}$ are Hermite polynomials.

\begin{figure}[h]
\begin{centering}
\includegraphics[scale=0.62]{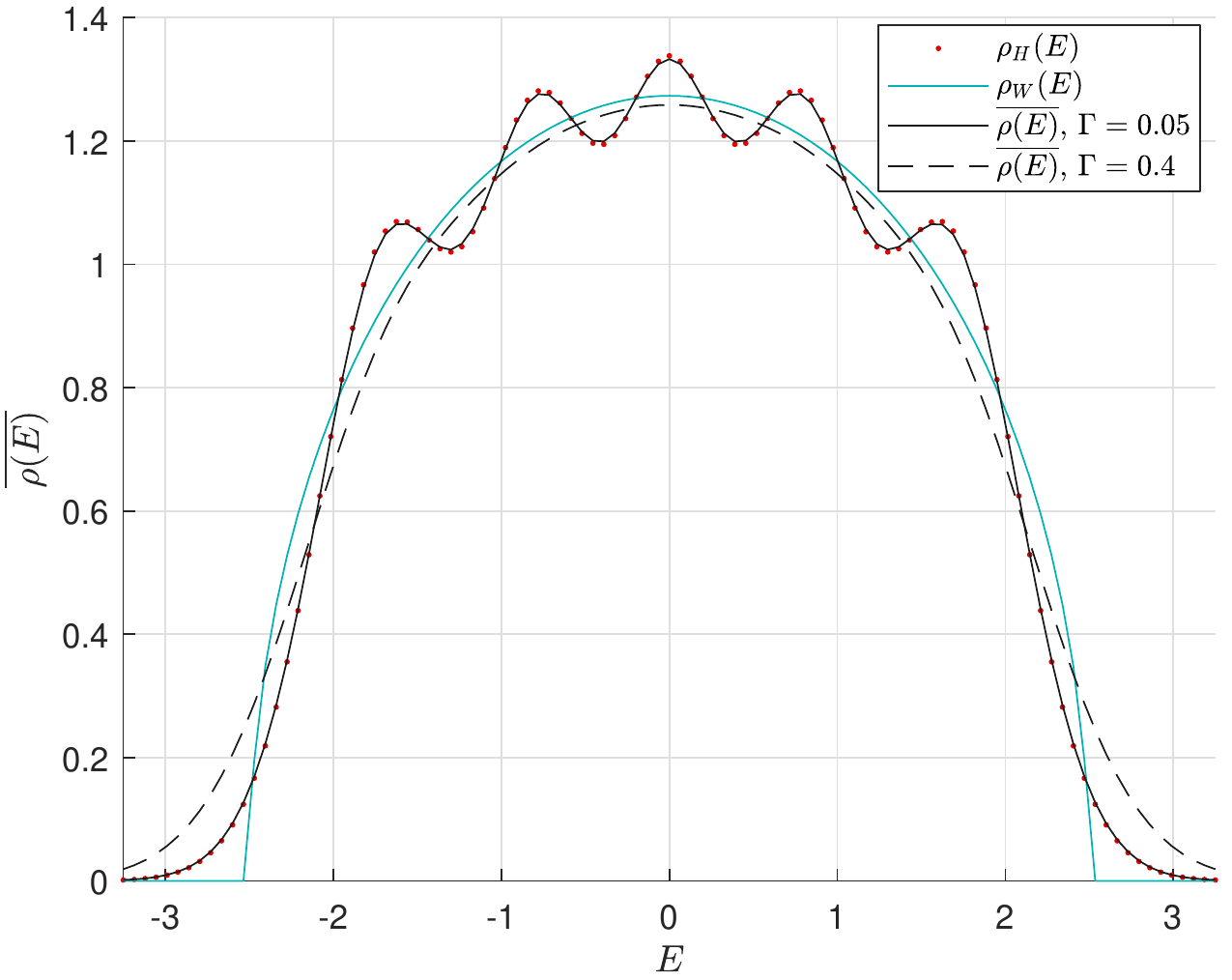}
\par\end{centering}
\caption{\label{den_2d}. Plot of the GUE level density, $\rho(E)$. The matrix
size is $N=5$ so that $a=2.5$. The solid black and dashed black
curves are $\overline{\rho(E)}$, the ensemble averaged level density
calculated using Eqs. (\ref{rho def}) and (\ref{delta gaussian}),
for $\Gamma=0.05$ and $\Gamma=0.4$ respectively. The ensemble size
was $k_{\mathrm{max}}=18\times10^{7}$. The solid cyan curve shows
$\rho_{W}(E)$ (Eq. \ref{semi}) and the red dots show $\rho_{H}(E)$
(Eq. (\ref{denH})).}
\end{figure}

In order to perform numerical calculations of the level density using
Eq. (\ref{rho def}), the delta function by may be represented \cite{Gomez2002,Sargeant2000,Feshbach1992}
by a Gaussian

\begin{equation}
\delta(E-E_{i})=\frac{1}{\sqrt{2\pi}\Gamma}\exp\Bigl(-\frac{(E-E_{i})^{2}}{2\Gamma^{2}}\Bigr),\label{delta gaussian}
\end{equation}
a Lorentzian

\begin{equation}
\delta(E-E_{i})=\frac{1}{\pi}\frac{\Gamma/2}{\left(E-E_{i}\right)^{2}+\Gamma^{2}/4},\label{eq:lorentzian}
\end{equation}
or a box function 

\begin{equation}
\delta(E-E_{i})=\frac{1}{\Gamma}\Bigl(\theta(E-E_{i}+\frac{\Gamma}{2})-\theta(E-E_{i}-\frac{\Gamma}{2})\Bigr).\label{eq:box}
\end{equation}
The width $\Gamma$ is chosen to be large enough to make neighbouring
levels overlap sufficiently to produce a smoothly varying density.
The Gaussian representation of the delta function was employed in
all the numerical simulations presented in this paper, since for a
given width, smooth level density, correlation and cluster functions
are obtained for smaller ensemble sizes than when the Lorentzian or
box function representations are employed. 

Fig. \ref{den_2d} compares numerical calculations of $\overline{\rho(E)}$,
the ensemble averaged level density calculated using Eqs. (\ref{rho def})
and (\ref{delta gaussian}), with $\rho_{W}(E)$ (Eq. \ref{semi})
and $\rho_{H}(E)$ (Eq. (\ref{denH})) for matrix size $N=5$. The
variance of the matrix elements in Eq. (\ref{probH}) is chosen such
that the radius of the Wigner semicircle is $a=N/2$, that is, we
choose $\sigma^{2}=\frac{N}{8\beta}=\frac{N}{16}$ or equivalently
$A=\frac{4\beta}{N}=\frac{8}{N}$. With this choice, the average level
density at the centre of the spectrum is given approximately by $\overline{\rho(0)}\approx\rho_{W}(0)=\frac{4}{\pi}\approx1.27$
for any $N$, a little higher for odd $N$, a little lower for even
$N$. 

The formula $\rho_{H}(E)$ for the average level density has $N$
oscillation peaks which disappear as $N$ goes to infinity resulting
in the Wigner semicircle $\rho_{W}(E)$. In Fig. \ref{den_2d} we
see that the oscillations are reproduced by the numerical simulations
for a Gaussian broadened delta function width as large as $\Gamma=0.05$
and that by the time the width is increased to $\Gamma=0.4$ the oscillations
are washed out. It can also be seen that as $\Gamma$ is increased,
$\overline{\rho(E)}$ lowers in the centre and rises at the edges,
as the functional form used to represent the delta function (Eq. (\ref{delta gaussian}))
starts to become visible. In Section \ref{sec: numerical}, $\Gamma$
dependence of the two-point correlation and cluster functions is observed
which corresponds to the $\Gamma$ dependence of $\overline{\rho(E)}$
seen here. 

\section{Two-point correlation and cluster functions: definitions and known
analytical results\label{sec:Two-point}}

A measure of how $\rho(E_{x}$) and $\rho(E_{y})$, the level density
at energies $E_{x}$ and $E_{y}$, are related is given by the two-point
correlation function \cite{PishroNik}:
\begin{align}
\rho_{2}(E_{x},E_{y}) & =\overline{\rho(E_{x})\rho(E_{y})}\label{eq:rho2}
\end{align}
Using Eq. (\ref{rho def}) for the level density in terms of the delta
function the correlation function may be written \cite{Fyodorov2005}

\begin{align}
\rho_{2}(E_{x},E_{y}) & =\sum_{i,j=1}^{N}\overline{\delta(E_{x}-E_{i})\delta(E_{y}-E_{j})}\nonumber \\
 & =\delta(E_{x}-E_{y})\sum_{i=1}^{N}\overline{\delta(E_{x}-E_{i})}\nonumber \\
 & +\sum_{i\ne j=1}^{N}\overline{\delta(E_{x}-E_{i})\delta(E_{y}-E_{j})}\nonumber \\
 & =\delta(E_{x}-E_{y})\overline{\rho(E_{x})}\nonumber \\
 & +\sum_{i\ne j=1}^{N}\overline{\delta(E_{x}-E_{i})\delta(E_{y}-E_{j})}\label{eq:rho2 separate}
\end{align}
Another definition of the two-point correlation function which measures
the probability density of finding a level at $E_{x}$ and a level
at $E_{y}$ while the position of the remaining levels is unobserved,
is given by \cite{Fyodorov2005}

\begin{align}
R_{2}(E_{x},E_{y}) & =\rho_{2}(E_{x},E_{y})-\delta(E_{x}-E_{y})\overline{\rho(E_{x})}\nonumber \\
 & =\sum_{i\ne j=1}^{N}\overline{\delta(E_{x}-E_{i})\delta(E_{y}-E_{j})}.\label{R2def}
\end{align}
The two-level cluster function \cite{Dyson1962} (essentially, the
negative of the autocovariance function \cite{PishroNik,Brody1981})
is given by

\begin{align}
T_{2}(E_{x},E_{y}) & =\overline{\rho(E_{x})}\ \overline{\rho(E_{y})}-R_{2}(E_{x},E_{y})\nonumber \\
 & =\sum_{i=1}^{N}\overline{\delta(E_{x}-E_{i})}\sum_{i=1}^{N}\overline{\delta(E_{y}-E_{i})}\nonumber \\
 & -\sum_{i\ne j=1}^{N}\overline{\delta(E_{x}-E_{i})\delta(E_{y}-E_{j})}.\label{T2def}
\end{align}
Wigner showed that for finite $N$, the two-point GUE cluster function
is given analytically by \cite{Wigner1962,Mehta1963,Mehta1967}

\begin{equation}
T_{H2}(E_{x},E_{y})=\exp\Bigl(-\frac{E_{x}^{2}+E_{y}^{2}}{2\sigma^{2}}\Bigr)\Bigl(\sum_{j=0}^{N-1}\pi_{j}(E_{x})\pi_{j}(E_{y})\Bigr){}^{2},\label{TH2}
\end{equation}
with the $\pi_{j}$ given by Eq. (\ref{piOrtho}). 

Brezin and Zee obtained an expression for the two-point GUE cluster
function which is valid for $\left|E_{y}-E_{x}\right|$ small enough
(and which is valid for more general probability distributions than
Eq. (\ref{probH}) as long as $\rho_{W}(E)$ is replaced correspondingly)
\cite{Brezin1993,Brezin1995}:

\begin{equation}
T_{S2}(E_{x},E_{y})=\rho_{W}(E_{x})\rho_{W}(E_{y})Y_{2}(E_{r})\label{TS2}
\end{equation}
where

\begin{equation}
E_{r}=\left(E_{x}-E_{y}\right)\rho_{W}((E_{x}+E_{y})/2)\label{eq:Er}
\end{equation}
and

\begin{equation}
Y_{2}(E_{r})=\frac{\sin^{2}\left(\pi E_{r}\right)}{\pi^{2}E_{r}^{2}}.\label{Y2}
\end{equation}
Here, $Y_{2}(x-y)$ is Dyson's expression for the two-point cluster
function which is valid for unfolded GUE levels, that is, for GUE
levels rescaled to constant unit average level density \cite{Dyson1962}.
(By $x$ and $y$ we mean two points on the unfolded energy scale.)

Another analytical expression for the $T_{2}(E_{x},E_{y})$, valid
for $\left|E_{y}-E_{x}\right|$ large enough and for \textbf{$\beta=1$},
2 and 4, is given by 

\begin{equation}
T_{L2}(E_{x},E_{y})=\frac{1}{\beta\pi^{2}}\frac{a^{2}-E_{x}E_{y}}{(E_{x}-E_{y})^{2}\sqrt{(a^{2}-E_{x}^{2})(a^{2}-E_{y}^{2})}}.\label{TL2}
\end{equation}
Eq. (\ref{TL2}) was obtained for the GOE ($\beta=1$) by French,
Mello and Pandey using the binary correlation method \cite{French1978}.
It has since been shown to be valid for the GUE ($\beta=2$) , GSE
($\beta=4$), and for probability distributions more general than
Eq. (\ref{probH}), using a variety of techniques which have their
origin in a variety of contexts \cite{Beenakker1994,Ambjorn1990,Brezin1993,Brezin1995,Khorunzhy1996,Pandey1981,Macedo1997,He2020}. 

Eq. (\ref{TL2}) does not describe the oscillations which are present
in Eqs. (\ref{TH2}) and (\ref{TS2}), but rather a smooth average
over the oscillations. In Ref. \cite{Khorunzhy1996} the smooth behaviour
is obtained analytically by maintaining the imaginary part of the
spectral parameter in a Stieltjes transform representation of the
cluster function large enough. In Refs. \cite{Brezin1993,Brezin1995}
the smooth behaviour is obtained analytically by replacing sines and
cosines by zero and their squares by 1/2 in oscillatory formulas.
In the numerical simulations presented in Section \ref{sec: numerical},
the smooth behaviour of Eq. (\ref{TL2}) is approached as $\Gamma$,
the width of the function representing the delta function, is increased. 

\section{Two-point correlation and cluster functions: numerical simulations\label{sec: numerical}}

In this Section, we present numerical simulations of the two-point
GUE correlation and cluster functions defined by Eqs. (\ref{R2def})
and (\ref{T2def}) with the delta function represented by Eq. (\ref{delta gaussian}).
Some details on how the ensemble average in Eq. (\ref{R2def}) was
coded are given in the Appendix. 

As in Section \ref{sec: level den}, the variance of the matrix elements
in Eq. (\ref{probH}) is chosen such that the radius of the Wigner
semicircle is $a=N/2$. With this choice, for any $N$, $T_{2}(0,0)\approx\left(\rho_{W}(0)\right)^{2}=\left(\frac{4}{\pi}\right)^{2}\approx1.62$
(the exact value being a little higher for odd $N$ and a little lower
for even $N$) and the zeros of $T_{2}(E_{x},E_{y})$ are similarly
spaced to the zeros of the unfolded cluster function $Y_{2}(x-y)$,
Eq. (\ref{Y2}).

\begin{figure}[h]
\begin{centering}
\includegraphics[scale=0.62]{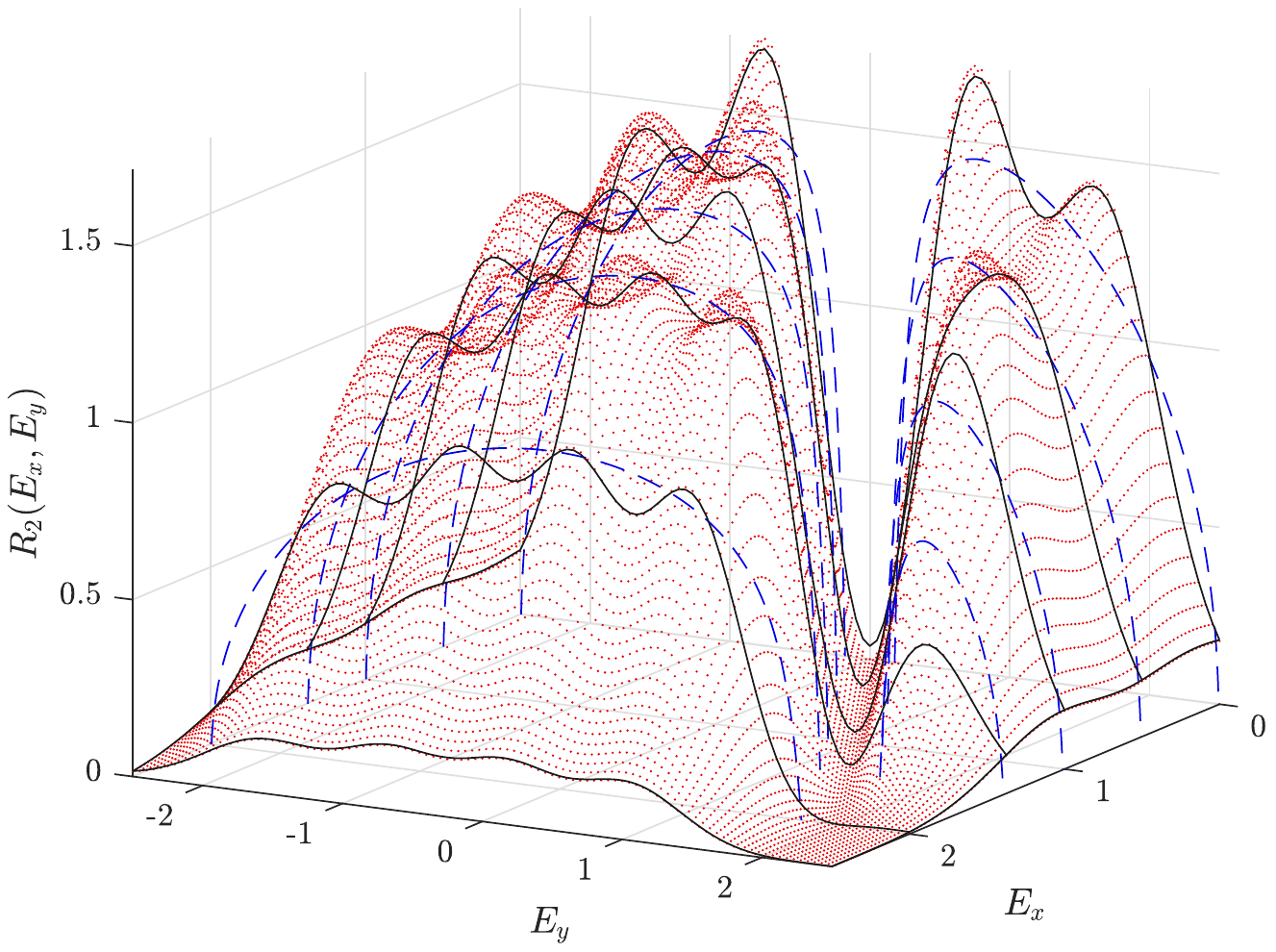}
\par\end{centering}
\caption{\label{R2}Plot of the two-point GUE correlation function $R_{2}(E_{x},E_{y})$.
The matrix size is $N=5$ so that $a=2.5$. The numerical simulations
(solid black curves) and $R_{L2}(E_{x},E_{y})=\rho_{W}(E_{x})\rho_{W}(E_{y})-T_{L2}(E_{x},E_{y})$
(dashed blue curves, see Eqs. (\ref{semi}) and (\ref{TL2})) are
shown along the lines $E_{x}=0$, $0.2a$, $0.4a$, 0.55$a$, 0.8$a$
and 0.9999$a$ and $E_{y}=-0.9999a$ and $0.9999a$. The numerical
simulations were carried out using Eqs. (\ref{R2def}) and (\ref{delta gaussian})
with Gaussian broadened delta function width $\Gamma=0.05$ and ensemble
size $k_{\mathrm{max}}=18\times10^{7}$. The red dots show $R_{H2}(E_{x},E_{y})=\rho_{H}(E_{x})\rho_{H}(E_{y})-T_{H2}(E_{x},E_{y})$
(see Eqs. (\ref{denH}) and (\ref{TH2})).}
\end{figure}

Fig. \ref{R2} displays calculations of the two-point GUE correlation
function for $N=5$. The solid black curves show the numerical simulations
of $R_{2}(E_{x},E_{y})$ as a function of $E_{y}$ for $E_{x}=0$,
$0.2a$, $0.4a$, 0.55$a$, 0.8$a$ and 0.9999$a$ and as a function
of $E_{x}$ for $E_{y}=-0.9999a$ and $0.9999a$. The numerical simulations
were carried out using Eqs. (\ref{R2def}) and (\ref{delta gaussian})
with delta function width $\Gamma=0.05$ and ensemble size $k_{\mathrm{max}}=18\times10^{7}$.
The dashed blue curves show the approximation $R_{L2}(E_{x},E_{y})=\rho_{W}(E_{x})\rho_{W}(E_{y})-T_{L2}(E_{x},E_{y})$
(see Eqs. (\ref{semi}) and (\ref{TL2})) for the same values of $E_{x}$
and $E_{y}$ as the black curves. The red dots show $R_{H2}(E_{x},E_{y})=\rho_{H}(E_{x})\rho_{H}(E_{y})-T_{H2}(E_{x},E_{y})$,
the exact two-point GUE correlation function in terms of Hermite polynomials
(see Eqs. (\ref{denH}) and (\ref{TH2})). The valley along $E_{x}=E_{y}$
corresponds to the well-known level repulsion of neighbouring levels.
Correlations involving levels at the edges are also seen to be small
but not zero and are smallest when both levels are near the same edge,
that is, near $E_{x}=E_{y}=a$. Correlations between the edge and
the centre the of spectrum, say between levels at $E_{x}=a$ and $E_{y}=0$
are slightly larger while correlations between opposite edges of the
spectrum, say between levels at $E_{x}=a$ and $E_{y}=-a$ are smaller
again, though not as small as correlations between levels near the
same edge. The numerical simulations are seen to correctly reproduce
the exact correlation function $R_{H2}(E_{x},E_{y})$. The approximation
$R_{L2}(E_{x},E_{y})$ is seen to produce a smooth average behaviour
over the oscillations except near the line $E_{x}=E_{y}$ and at the
edges, where $T_{L2}(E_{x},E_{y})$ is singular.

\begin{figure}[h]
\begin{centering}
\includegraphics[scale=0.62]{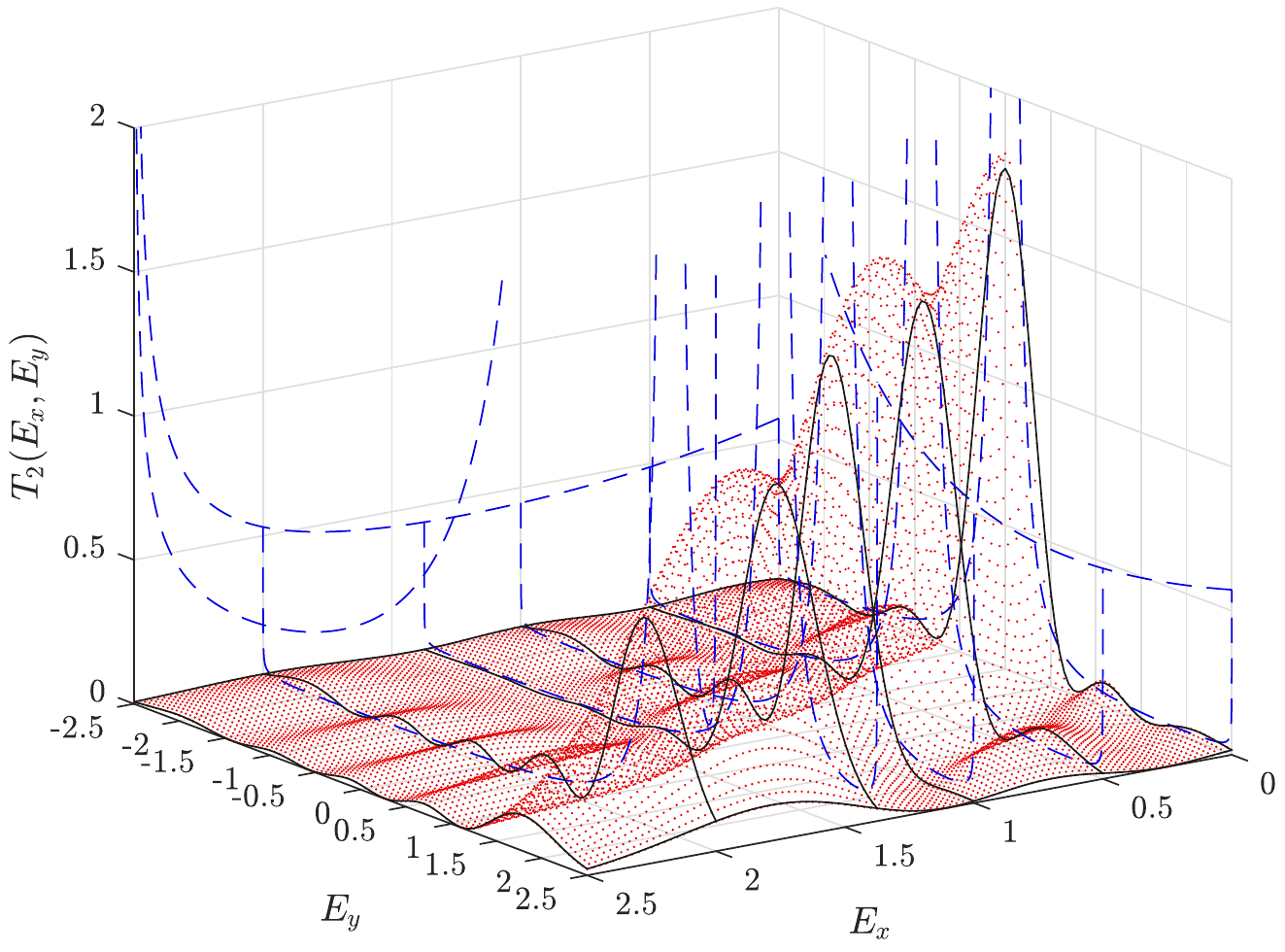}
\par\end{centering}
\caption{\label{T2}Plot of the two-point GUE cluster function $T_{2}(E_{x},E_{y})$.
The matrix size is $N=5$ so that $a=2.5$. The numerical simulations
(solid black curves) and $T_{L2}(E_{x},E_{y})$ (dashed blue curves,
see Eq. (\ref{TL2})) are shown for the same values of $E_{x}$ and
$E_{y}$ as in Fig. \ref{R2}. The numerical simulations were carried
out using Eqs. (\ref{T2def}) and (\ref{delta gaussian}) for the
same values of $\Gamma$ and $k_{\mathrm{max}}$ as Fig. \ref{R2}.
The red dots show $T_{H2}(E_{x},E_{y})$ (Eq. (\ref{TH2})).}
\end{figure}

Fig. \ref{T2} displays calculations of the two-point GUE cluster
function for $N=5$. Corresponding to the $N$ maxima of the level
density (see Fig. \ref{den_2d}) there are $N$ maxima along the line
$E_{x}=E_{y}$ which reduce in height with increasing $E_{x}$ and
$E_{y}$. (The region $E_{x}<0$ is not shown since $T_{2}(E_{x},E_{y})=T(E_{y},E_{x})$).)
The numerical simulations of $T_{2}(E_{x},E_{y})$ (solid black curves)
were carried out using Eqs. (\ref{T2def}) and (\ref{delta gaussian})
for the same values of $\Gamma$ and $k_{\mathrm{max}}$ as Fig. \ref{R2}.
The numerical simulations and the approximation $T_{L2}(E_{x},E_{y})$
(dashed blue curves, see Eq. (\ref{TL2})) are shown for the same
values of $E_{x}$ and $E_{y}$ as the solid black and dashed blue
curves in Fig. \ref{R2}. The red dots show $T_{H2}(E_{x},E_{y})$
, the exact two-point GUE cluster function in terms of Hermite polynomials
(Eq. (\ref{TH2})). Similarly to correlation function, the numerical
simulations are seen to correctly reproduce the exact cluster function
$T_{H2}(E_{x},E_{y})$. Again, the approximation $T_{L2}(E_{x},E_{y})$
is seen to produce smooth average behaviour over the oscillations
except near the line $E_{x}=E_{y}$ and at the edges, where $T_{L2}(E_{x},E_{y})$
is singular.

\begin{figure}[h]
\begin{centering}
\includegraphics[scale=0.62]{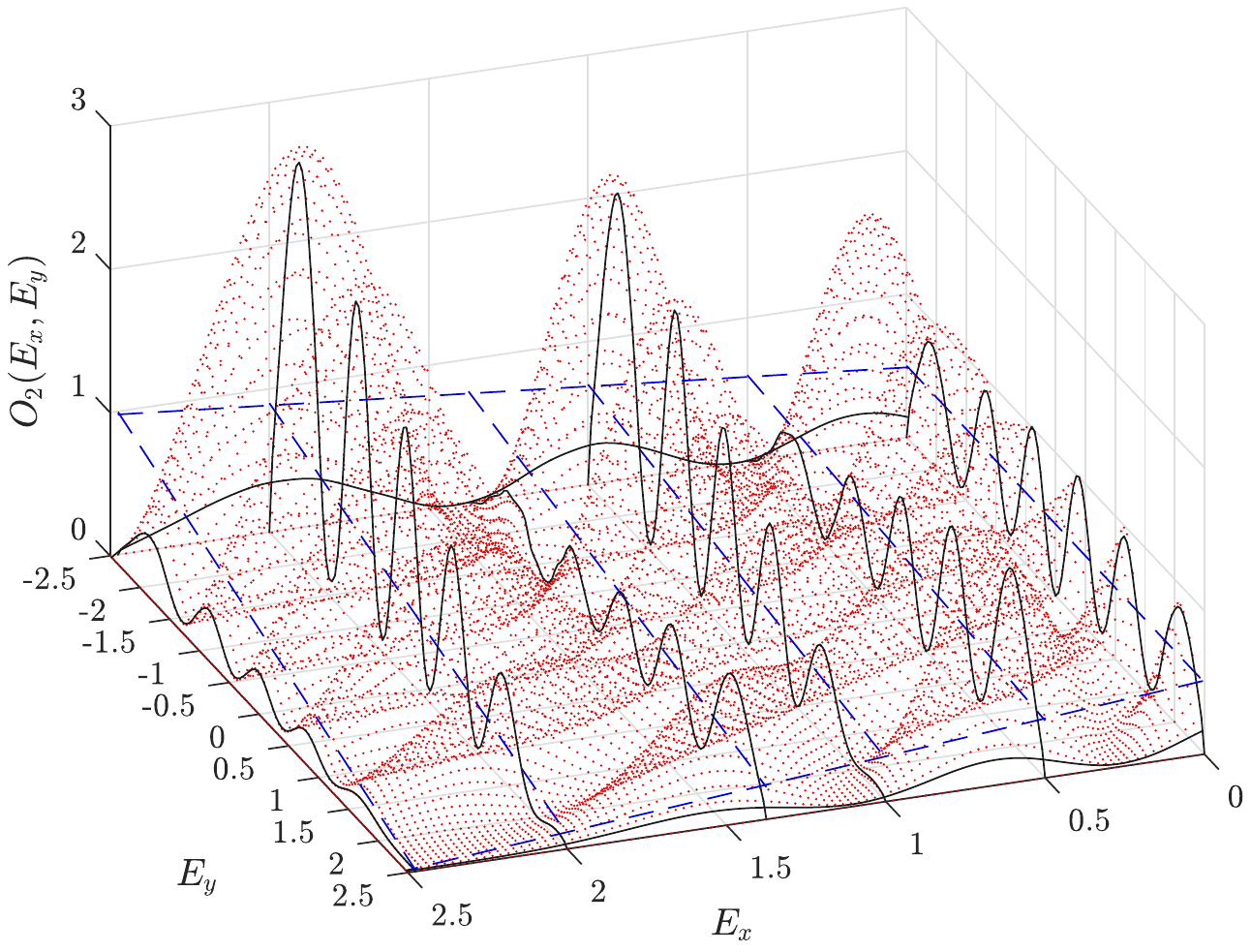}
\par\end{centering}
\caption{\label{O2}Plot of the function $O_{2}(E_{x},E_{y})$ for the GUE.
The matrix size is $N=5$ so that $a=2.5$. The numerical simulations
(solid black curves) and $O_{L2}(E_{x},E_{y})$ (dashed blue curves,
see Eq. (\ref{O2L})) are shown along the lines $E_{x}=0$, $0.2a$,
$0.4a$, $0.55a$, 0.8$a$ and 0.99$a$ and $E_{y}=-0.99a$ and $0.99a$.
The numerical simulations were carried out using Eqs. (\ref{O2def}),
(\ref{U2}), (\ref{T2def}) and (\ref{delta gaussian}) for the same
values of $\Gamma$ and $k_{\mathrm{max}}$ as Fig. \ref{R2}. The
red dots show $O_{H2}(E_{x},E_{y})=U_{2}(E_{x},E_{y})T_{H2}(E_{x},E_{y})$
(see Eqs. (\ref{U2}) and (\ref{TH2})).}
\end{figure}

The oscillations of $T_{2}(E_{x},E_{y})$ continue out to large values
of $\left|E_{y}-E_{x}\right|$ but are difficult to see in Fig. \ref{T2}
because of the rapid decay of $T_{2}(E_{x},E_{y})$ with increasing
$\left|E_{y}-E_{x}\right|$. To isolate the oscillatory behaviour,
in Fig. \ref{O2} we plot the function 

\begin{equation}
O_{2}(E_{x},E_{y})=U_{2}(E_{x},E_{y})T_{2}(E_{x},E_{y}),\label{O2def}
\end{equation}
where
\begin{equation}
U_{2}(E_{x},E_{y})=\frac{\beta\pi^{2}}{2a^{2}}(E_{x}-E_{y})^{2}\sqrt{(a^{2}-E_{x}^{2})(a^{2}-E_{y}^{2})}.\label{U2}
\end{equation}
Here, $U_{2}(E_{x},E_{y})$ is essentially the denominator of $T_{L2}(E_{x},E_{y})$.
Multiplying Eq. (\ref{TL2}) by $U_{2}(E_{x},E_{y})$ we obtain

\begin{equation}
O_{L2}(E_{x},E_{y})=\frac{1}{2}\Bigl(1-\frac{E_{x}E_{y}}{a^{2}}\Bigr).\label{O2L}
\end{equation}
The solid black curves in Fig. \ref{O2} show numerical simulations
of $O_{2}(E_{x},E_{y})$ for $N=5$ carried out using Eqs. (\ref{O2def}),
(\ref{U2}), (\ref{T2def}) and (\ref{delta gaussian}) for the same
values of $\Gamma$ and $k_{\mathrm{max}}$ as were used in Fig. \ref{R2}
and \ref{T2}. The numerical simulations and $O_{L2}(E_{x},E_{y})$
(dashed blue curves, see Eq. (\ref{O2L})) are shown along the lines
$E_{x}=0$, $0.2a$, $0.4a$, $0.55a$, 0.8$a$ and 0.99$a$ and $E_{y}=-0.99a$
and $0.99a$. The red dots show $O_{H2}(E_{x},E_{y})=U_{2}(E_{x},E_{y})T_{H2}(E_{x},E_{y})$,
the exact expression in terms of Hermite polynomials (see Eqs. (\ref{U2})
and (\ref{TH2})). The oscillations are visible for all values of
$E_{x}$ and $E_{y}$ and the numerical simulations are seen to correctly
reproduce the oscillations of the exact expression $O_{H2}(E_{x},E_{y})$.
In can be seen that $O_{2}(0,E_{y})$ oscillates around an average
value of $O_{L2}(0,E_{y})=\frac{1}{2}$ between $E_{y}=-a$ and $a$.
However, since $O_{2}(E_{x},E_{y})$ oscillates in both the $E_{x}$
and $E_{y}$ directions there are valleys visible along which $O_{L2}(E_{x},E_{y})$
overestimates $O_{2}(E_{x},E_{y})$. In particular, $O_{L2}(E_{x},E_{y})$
overestimates $O_{2}(E_{x},E_{y})$ near the line $E_{x}=E_{y}$ and
near the edges, where $T_{L2}(E_{x},E_{y})$ is singular. Previously,
Kobayakawa et al. \cite{Kobayakawa1995} numerically investigated
$-(E_{x}-E_{y})^{2}T_{2}(E_{x},E_{y})$ rather than $O_{2}(E_{x},E_{y})$
for generalisations of the GUE, GOE and GSE. In particular, these
authors were interested in verifying numerically that the probability
distribution of the matrix elements and corresponding level density
only appear in the smoothed cluster function, Eq. (\ref{TL2}), through
the parameter $a$ which characterises the width of the spectrum. 

\begin{figure}[h]
\begin{centering}
\includegraphics[scale=0.62]{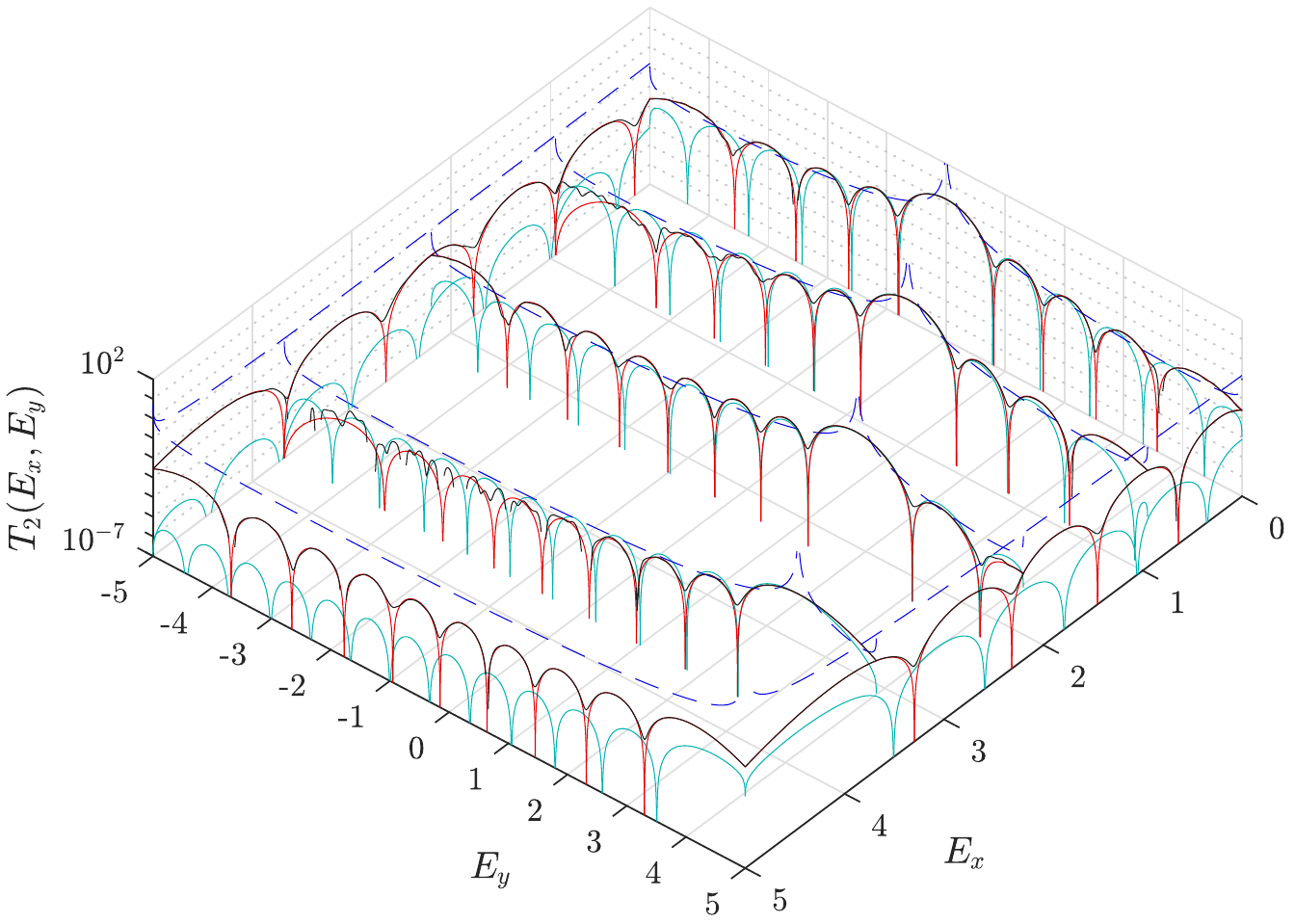}
\par\end{centering}
\caption{\label{logT2N10Gam0.05}Semi-logarithmic plot of the two-point GUE
cluster function $T_{2}(E_{x},E_{y})$ along the lines $E_{x}=0$,
$0.19a$, 0.44$a$, 0.736$a$ and 0.9999$a$ and $E_{y}=-0.9999a$
and $0.9999a$. The matrix size is $N=10$ so that $a=5$. The solid
black curves show the numerical simulations carried out using Eqs.
(\ref{T2def}) and (\ref{delta gaussian}) with delta function width
$\Gamma=0.05$ and ensemble size $k_{\mathrm{max}}=9\times10^{7}$.
The red solid curves show $T_{H2}(E_{x},E_{y})$ (Eq. (\ref{TH2})),
the solid cyan curves show $T_{S2}(E_{x},E_{y})$ (Eq. (\ref{TS2}))
and the dashed blue curves show $T_{L2}(E_{x},E_{y})$ (Eq. (\ref{TL2})).}
\end{figure}

\begin{figure}[h]
\begin{centering}
\-\includegraphics[scale=0.62]{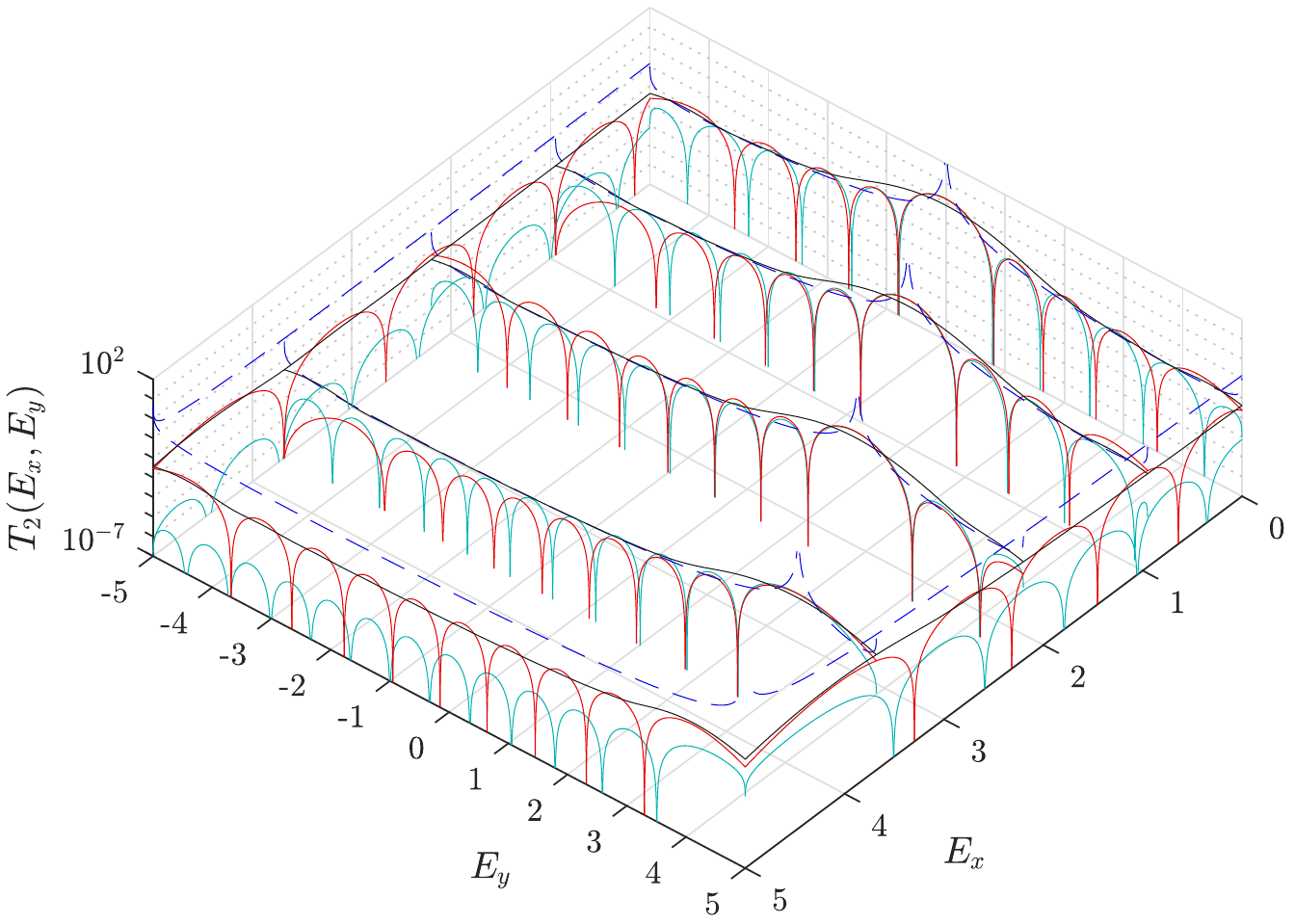}
\par\end{centering}
\caption{\label{logT2N10Gam0.4}Same as Fig. (\ref{logT2N10Gam0.05}) but with
$\Gamma=0.4$.}
\end{figure}

Figs. \ref{logT2N10Gam0.05} and \ref{logT2N10Gam0.4} display semi-logarithmic
plots of the two-point GUE cluster function $T_{2}(E_{x},E_{y})$
for $N=10$ along the lines $E_{x}=0$, $0.19a$, 0.44$a$, 0.736$a$
and 0.9999$a$ and $E_{y}=-0.9999a$ and $0.9999a$. The red solid
curves show the exact result $T_{H2}(E_{x},E_{y})$ (Eq. (\ref{TH2})),
the solid cyan curves show the approximation $T_{S2}(E_{x},E_{y})$
(Eq. (\ref{TS2})) and the dashed blue curves show the approximation
$T_{L2}(E_{x},E_{y})$ (Eq. (\ref{TL2})). The solid black curves
show the numerical simulations carried out using Eqs. (\ref{T2def})
and (\ref{delta gaussian}) for ensemble size $k_{\mathrm{max}}=9\times10^{7}$
with the delta function width $\Gamma=0.05$ in Fig. \ref{logT2N10Gam0.05}
and $\Gamma=0.4$ in Fig. \ref{logT2N10Gam0.4}. It is seen that when
$\Gamma=0.05$ the numerical simulations follow the oscillations of
the exact result $T_{H2}(E_{x},E_{y})$, but when the delta function
width is increased to $\Gamma=0.4$ the numerical simulations follow
the smooth behaviour of $T_{L2}(E_{x},E_{y})$ for large enough $\left|E_{y}-E_{x}\right|$.
This transition corresponds to the transition which is observed in
Fig. \ref{den_2d} for the average level density as $\Gamma$ is increased:
the average level density changes from oscillating to smooth and lowers
in the centre while rising at the edges. We mention two numerical
artifacts which are visible in Figs. \ref{logT2N10Gam0.05} and \ref{logT2N10Gam0.4}
which are related to the fact the density of levels is larger at the
centre of the spectrum than the edges. First, in Fig. \ref{logT2N10Gam0.05}
the numerical simulations fail to correctly produce a smooth curve
in the regions where $T_{2}(E_{x},E_{y})$ is smallest. This occurs
because in these regions insufficient terms in the sums of delta functions
in Eq. (\ref{T2def}) make non-zero contributions; this could be fixed
by using a larger ensemble. Second, in Fig. \ref{logT2N10Gam0.4}
the peaks in the numerical simulations at $E_{x}=E_{y}$ are artificially
broad. This occurs be cause many more terms in the sums of delta functions
in Eq. (\ref{T2def}) make non-zero contributions in this region than
at the edges resulting in the functional form used to represent the
delta function (Eq. (\ref{delta gaussian})) becoming visible more
quickly at the centre than at the edges as the delta function width
$\Gamma$ is increased. It may be possible fix this by using a smaller
delta function width at the centre of the spectrum than at the edges
but we do not explore this possibility here. We also note that the
approximation $T_{S2}(E_{x},E_{y})$ is seen to correctly produce
the oscillations of the exact result $T_{H2}(E_{x},E_{y})$ for small
enough $\left|E_{y}-E_{x}\right|$ but to underestimate it near the
edges. 
\begin{figure}[h]
\begin{centering}
\includegraphics[scale=0.62]{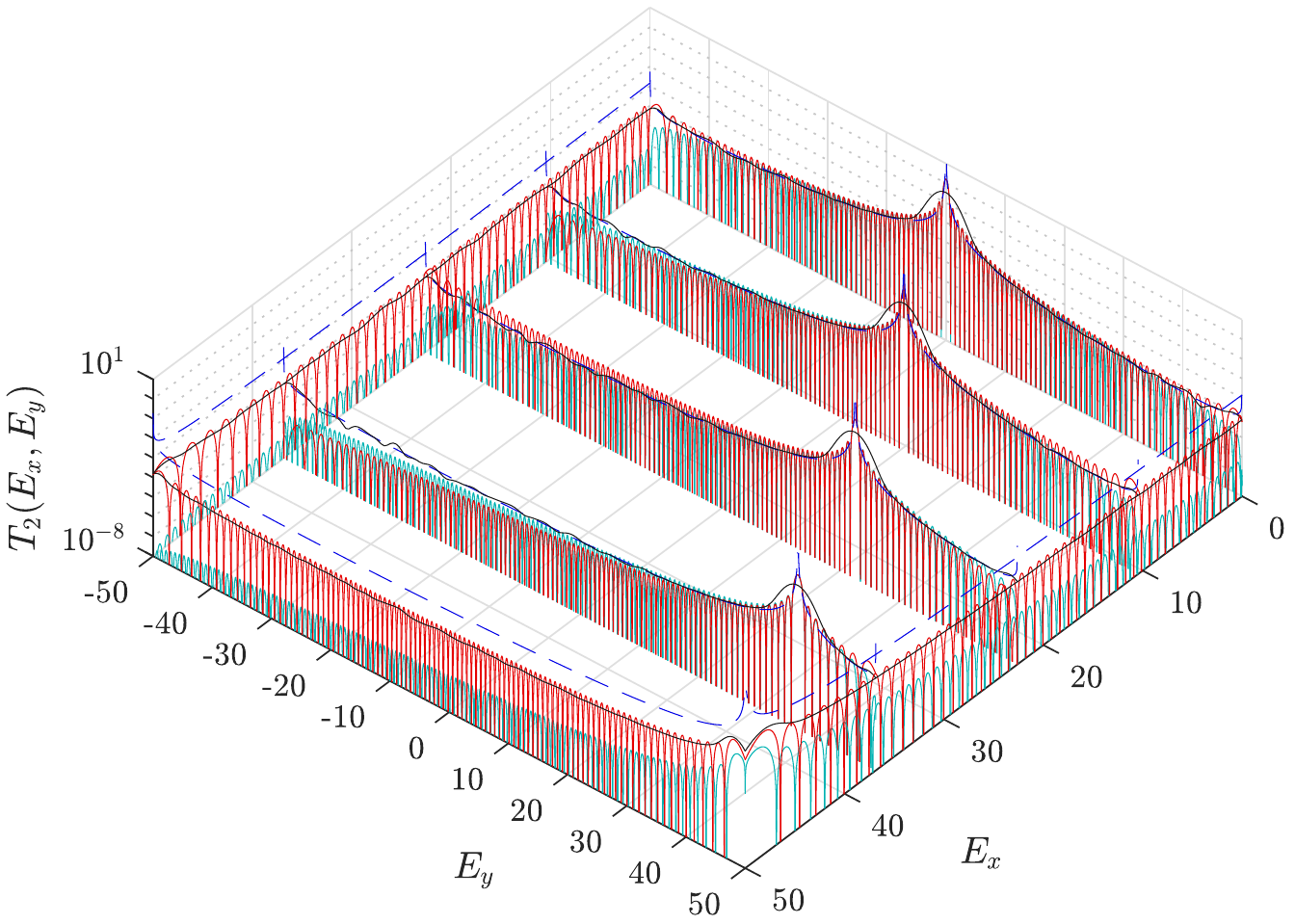}
\par\end{centering}
\caption{\label{logT2N100}Semi-logarithmic plot of the two-point GUE cluster
function $T_{2}(E_{x},E_{y})$ along the lines $E_{x}=0$, $0.2093a$,
0.452$a$, 0.7372$a$ and 0.9999$a$ and $E_{y}=-0.9999a$ and $0.9999a$.
The matrix size is $N=100$ so that $a=50$. The meaning of the line
styles is the same as Fig. \ref{logT2N10Gam0.05}. The numerical simulations
were carried out for $\Gamma=1$ and $k_{\mathrm{max}}=9\times10^{6}$.}
\end{figure}

Figs. \ref{logT2N100} displays semi-logarithmic plots of the two-point
GUE cluster function $T_{2}(E_{x},E_{y})$ for $N=100$ along the
lines $E_{x}=0$, $0.2093a$, 0.452$a$, 0.7372$a$ and 0.9999$a$
and $E_{y}=-0.9999a$ and $0.9999a$. The numerical simulations were
carried out for $k_{\mathrm{max}}=9\times10^{6}$ and $\Gamma=1$.
For this value of $\Gamma$ the numerical simulations follow the smooth
behaviour of $T_{L2}(E_{x},E_{y})$ rather than the oscillatory behaviour
of $T_{H2}(E_{x},E_{y})$. To correctly reproduce the oscillatory
behaviour of $T_{H2}(E_{x},E_{y})$ with numerical simulations it
is necessary to use a Gaussian broadened delta function width of no
more than $\Gamma\sim1/20$ while to obtain a smoothly varying cluster
function for this order of $\Gamma$ it is necessary to use an ensemble
of eigenvalues of size $k_{\mathrm{max}}\sim10^{8}$. For $N=100$,
an ensemble of eigenvalues of size $k_{\mathrm{max}}=10^{8}$ occupies
80 GB of computer memory and we were unable to perform the corresponding
ensemble averages in a reasonable time with the computer which was
available (9th generation Intel Core i9 with 64 GB of RAM). Again,
the approximation $T_{S2}(E_{x},E_{y})$ is seen to correctly produce
the oscillations of the exact result $T_{H2}(E_{x},E_{y})$ for small
enough $\left|E_{y}-E_{x}\right|$ but to underestimate it near the
edges.

\begin{figure}[h]
\begin{centering}
\includegraphics[scale=0.62]{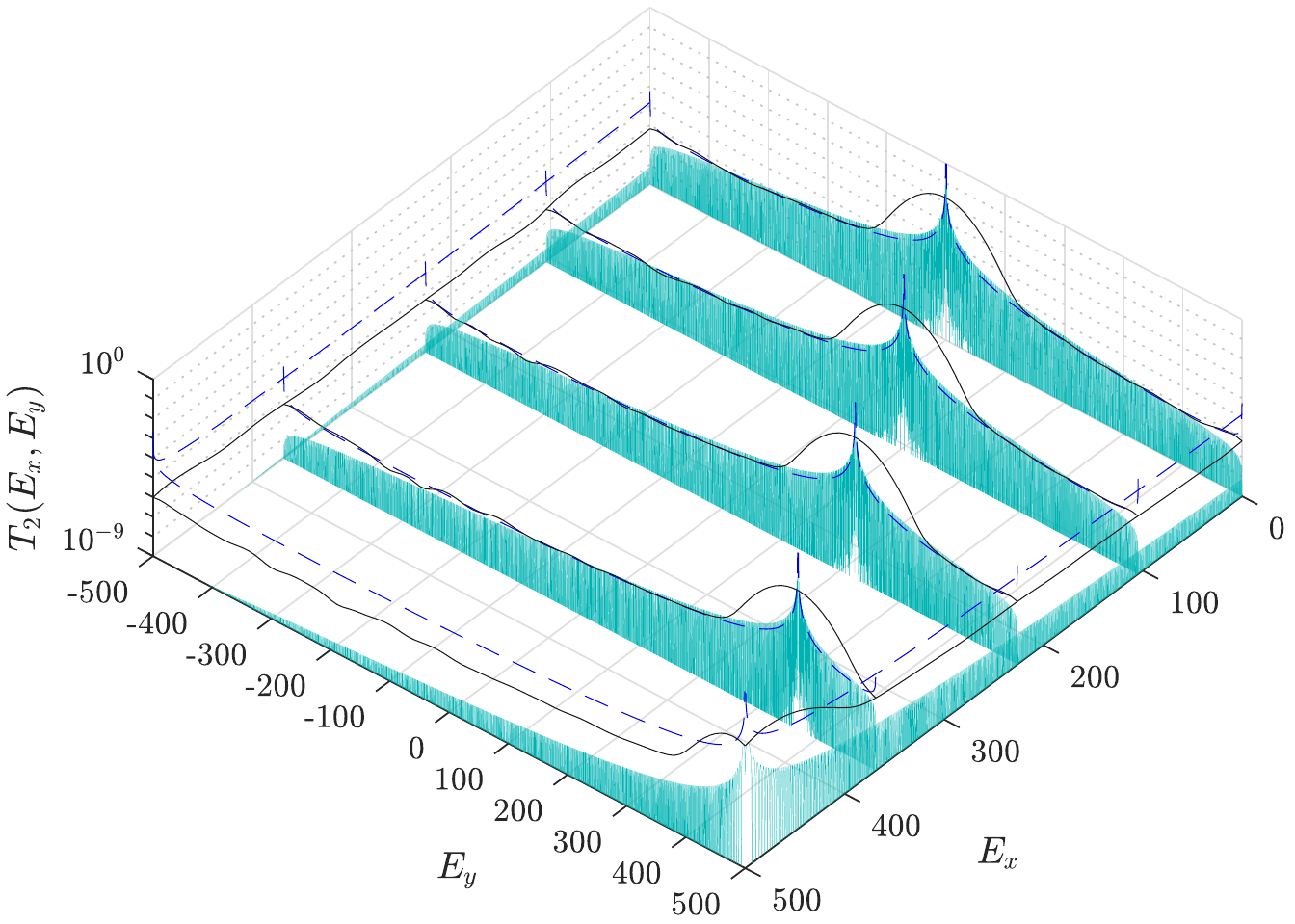}
\par\end{centering}
\caption{\label{logT2N1000}Semi-logarithmic plot of the two-point GUE cluster
function $T_{2}(E_{x},E_{y})$ for the same values of $E_{x}$ and
$E_{y}$ as a fraction of $a$ as Fig. \ref{logT2N100}. The matrix
size is $N=1000$ so that $a=500$. The meaning of the line styles
is the same as Fig. \ref{logT2N10Gam0.05}. (Eq. (\ref{TH2}) does
work numerically for $N=1000$ so no plot of $T_{H2}(E_{x},E_{y})$
is included is this graph.) The numerical simulations were carried
out for $\Gamma=20$ and $k_{\mathrm{max}}=9\times10^{5}$.}
\end{figure}

Fig. \ref{logT2N1000} displays semi-logarithmic plots of the two-point
GUE cluster function $T_{2}(E_{x},E_{y})$ for $N=1000$ for the same
values of $E_{x}$ and $E_{y}$ as a fraction of $a$ as Fig. \ref{logT2N100}.
The numerical simulations were carried out for $k_{\mathrm{max}}=9\times10^{5}$
and $\Gamma=20$. For this value of $\Gamma$ the numerical simulations
again follow the smooth behaviour of $T_{L2}(E_{x},E_{y})$ rather
than the oscillatory behaviour of $T_{H2}(E_{x},E_{y})$. The exact
expression for $T_{H2}(E_{x},E_{y})$ in terms of Hermite polynomials
(Eq. (\ref{TH2})) does work numerically for $N=1000$ so no plot
of $T_{H2}(E_{x},E_{y})$ is included in Fig. \ref{logT2N1000}. The
approximation $T_{S2}(E_{x},E_{y})$ is shown for comparison and can
be seen to oscillate around $T_{L2}(E_{x},E_{y})$ except near the
edges. In Figs. \ref{logT2N10Gam0.05}, \ref{logT2N10Gam0.4}, \ref{logT2N100}
and \ref{logT2N1000} it can seen that near the edges $T_{S2}(E_{x},E_{y})$
underestimates $T_{2}(E_{x},E_{y})$ while $T_{L2}(E_{x},E_{y})$
overestimates it. As explained in relation to Fig. \ref{logT2N10Gam0.4},
the artificially broad peaks which occur in the numerical simulations
at $E_{x}=E_{y}$ are due to the large value of delta function width
used ($\Gamma=20$). It was necessary to use such a large delta function
width to compensate for the relatively small ensemble size used ($k_{\mathrm{max}}=9\times10^{5}$).
Artificially broad peaks at $E_{x}=E_{y}$ are also visible to a lesser
extent in the numerical simulations of Fig. \ref{logT2N100} for which
$\Gamma=1$ and the ensemble size was $k_{\mathrm{max}}=9\times10^{6}$.

\section{Summary and conclusions\label{concl}}

Numerical simulations of the two-point eigenvalue correlation and
cluster functions of the Gaussian unitary ensemble were carried out
directly from their definitions by Eqs. (\ref{R2def}) and (\ref{T2def})
in terms of deltas functions. In Figs. \ref{R2} - \ref{logT2N1000},
the numerical simulations are compared with analytical results which
follow from three analytical formulas for the two-point GUE cluster
function: (i) Eq. (\ref{TH2}), (ii) Eq. (\ref{TS2}) and (iii) Eq.
(\ref{TL2}). It is found that the oscillations present in formulas
(i) and (ii) are reproduced by the numerical simulations if the the
width of the function used to represent the delta function is small
enough and that the non-oscillating behaviour of formula (iii) is
approached as the width is increased.

It should be possible to use this method of numerically simulating
two-point correlation and cluster functions to investigate for the
range of validity of analytical results for other ensembles such as
deformed ensembles \cite{Bertuola2005,Guhr1998} and to assist in
the investigation of ensembles which are currently intractable analytically. 
\begin{acknowledgements}
I would like to acknowledge happy collaborations with Mahir Saleh
Hussein and Mauricio Porto Pato which involved random matrices in
nuclear physics and were the precursor to this work. 

\setcounter{equation}{0} 
\renewcommand\theequation{A\arabic{equation}}
\section*{Appendix}
In this appendix we offer some considerations on the coding of the ensemble average in Eq. (\ref{R2def}) for the two-point correlation function. Let us denote the $i$th eigenvalue of the $k$th realisation of the ensemble by $E_{ik}$ and define  $N\times k_{\mathrm{max}}$ matrices $\verb!X!$ and $\verb!Y!$ whose elements are $\delta(E_x - E_{ik})$ and $\delta(E_y - E_{ik})$ respectively. Then the two-point correlation function $\verb!R!$ of energies $E_x$ and $E_y$ may be conveniently coded in Octave or Matlab by
\begin{eqnarray}
&&\verb!D = X * Y'!\\ 
&&\verb!R = (sum(D(:)) - trace(D))/kmax!
\end{eqnarray}
where the $N\times N$ matrix $\verb!D!$ is the outer product of $\verb!X!$ and $\verb!Y'!$ (the transpose of $\verb!Y!$) and $\verb!sum(D(:))!$ is the grand sum of its elements. 

\bibliographystyle{spphys}
\bibliography{gueCorrelationFunctions}
\end{acknowledgements}

\end{document}